\documentclass[12pt]{article}
\usepackage{epsfig}
\usepackage{a4wide}

\newcommand{\GeV}{\mbox{ GeV}} 
\newcommand{\TeV}{\mbox{ TeV}}
 
\newcommand{\sq}{\ensuremath{\tilde q}}

\newcommand{\stopp}{\ensuremath{\tilde t}} 
\newcommand{\sbottom}{\ensuremath{\tilde b}}  
\newcommand{\stau}{\ensuremath{\tilde \tau}}

\newcommand{\cplus}{\ensuremath{\chi^+}} 
\newcommand{\cmin}{\ensuremath{\chi^-}} 
\newcommand{\neut}{\ensuremath{\chi^0}}

\newcommand{\sfr}{{\tilde{f}}}

\newcommand{\tb}{\ensuremath{\tan\beta}}

\newcommand{\mb}{\ensuremath{m_b}} 
 
\newcommand{\mt}{\ensuremath{m_t}}

\newcommand{\mw}{\ensuremath{M_W}} 
\newcommand{\mws}{\ensuremath{M^2_W}}

\newcommand{\mxs}{\ensuremath{M^2_X}}
\newcommand{\mx}{\ensuremath{M_X}}

\newcommand{\mHp}{\ensuremath{M_{H^\pm}}} 
\newcommand{\mA}{\ensuremath{M_{A^0}}}

\newcommand{\mi}{\ensuremath{M_i}} 
\newcommand{\mis}{\ensuremath{M^2_i}}

\newcommand{\mg}{\ensuremath{m_{\tilde{g}}}}

\newcommand{\msbo}{\ensuremath{m_{\tilde{b}_1}}} 
 
\newcommand{\msbt}{\ensuremath{m_{\tilde{b}_2}}}

\newcommand{\msto}{\ensuremath{m_{\tilde{t}_1}}} 
\newcommand{\mstt}{\ensuremath{m_{\tilde{t}_2}}}

\newcommand{\msdo}{\ensuremath{m_{\tilde d_1}}} 
\newcommand{\msdt}{\ensuremath{m_{\tilde d_2}}} 
 
 \newcommand{\msut}{\ensuremath{m_{\tilde u_2}}}

\newcommand{\mselo}{\ensuremath{m_{\tilde{e}_1}}} 
\newcommand{\mselt}{\ensuremath{m_{\tilde{e}_2}}} 
 
\newcommand{\mwf}{{\mw^4}}

\newcommand{\cbta}{c_\beta}

\newcommand{\cfbt}{c_{4\beta}}
\newcommand{\sbta}{s_\beta}
\newcommand{\stbt}{s_{2\beta}}

\newcommand{\sws}{s_W^2}

\newcommand{\UChaio}{U_{i1}}
\newcommand{\UChait}{U_{i2}}

\newcommand{\mchasomenysmchast}{\left(M_1^2-M_2^2\right)}
\newcommand{\commonnumfactor}{\frac{\alpha}{4\,\pi\,\sws}}

\newcommand{\osf}{\ensuremath{\theta_f}}

\newcommand{\osb}{\ensuremath{\theta_b}}
\newcommand{\ost}{\ensuremath{\theta_t}}
\newcommand{\osd}{\ensuremath{\theta_d}}
\newcommand{\osu}{\ensuremath{\theta_u}}
\newcommand{\osel}{\ensuremath{\theta_e}}

\begin{document}

\vspace*{-1.5cm}
\begin{flushright}
{\parbox{10cm}{PSI-PR-03-09, LC-TH-2003-033, hep-ph/0307011}}
\end{flushright}

\begin{center}

{\Large \textbf{Electroweak radiative corrections to sfermion
    decays}\footnote{Updated talk given at the 2nd Workshop of the
    ``Extended Joint ECFA/DESY Study on Physics and 
 Detectors for a
  Linear Electron - Positron Collider'' Saint Malo (France), 12-15 April 2002. }
} 

\vspace{0.8cm}

{\large \underline{Jaume Guasch}$^{a}$,  
Wolfgang Hollik$^{b}$, Joan Sol\`a$^{c,d}$}

\vspace*{0.2cm}
{\sl 
$^a$ Theory Group LTP, Paul
    Scherrer Institut, CH-5232 Villigen PSI, Switzerland\\
$^b$  Max-Planck-Institut f\"ur Physik,
   F\"ohringer Ring 6, D-80805 M\"unchen, Germany \\
$^c$  Departament d'Estructura i Constituents de la 
Mat\`eria,  Universitat de Barcelona, Diagonal 647, E-08028 Barcelona,
  Catalonia, Spain\\
$^d$ Institut de F\'{\i}sica d'Altes Energies, Universitat Aut\`onoma de
  Barcelona, E-08193 Bellaterra, Barcelona, Catalonia, Spain} 

\end{center}


 \begin{abstract}
\noindent
We analyze the partial decay widths of sfermions decaying into charginos
and neutralinos $\Gamma(\sfr\to f'\chi)$ at the one-loop level,
including the electroweak and strong corrections. We
present the renormalization framework, and discuss the value of the
corrections. Since these corrections show non-decoupling effects, we
analyze the radiative effects induced by a heavy squark sector into the
lepton-slepton-chargino/neutralino couplings. We conclude that some
knowledge of the heavy sector is needed in order to provide a sufficiently
precise prediction for slepton observables at an $e^+e^-$ Linear Collider.
 \end{abstract}

\section{Introduction}

One of the basic predictions of Supersymmetry (SUSY) is the equality between
the couplings of SM particles and that of their superpartners. The
simplest processes in which this prediction could be tested is the partial
decay widths of sfermions into Standard Model (SM) fermions and
charginos/neutralinos:
\begin{equation}
\Gamma(\sfr \to f' \chi)\,\,.
\label{eq:gammadef}
\end{equation}
By measuring these partial decay widths (or the corresponding branching
ratios) one could measure the fermion-sfermion-chargino/neutralino
Yukawa couplings and compare them with the SM fermion gauge
couplings. The $e^+e^-$ Linear Collider (LC) is an ideal machine where
these test can be performed, with a precision below the percent level.

We have computed the full one-loop electroweak corrections to the
partial decay widths~(\ref{eq:gammadef}). As we will show, the radiative
corrections induce finite shifts in the couplings which are non-decoupling.

The QCD corrections to the process~(\ref{eq:gammadef}) were computed
in~\cite{QCD}, 
and the Yukawa corrections to
bottom-squarks decaying into charginos was given in~\cite{Guasch:1998as}.
Here we present the last step, namely, the full electroweak corrections
in the framework of the Minimal Supersymmetric Standard Model (MSSM).
Full details of the present work can be found
in~\cite{EWcorr}. The present note complements and supersedes Ref.~\cite{Guasch:LCnote}.

\section{Renormalization and radiative corrections}

The computation to one-loop level of the partial decay
width~(\ref{eq:gammadef}) requires the renormalization of the full MSSM
Lagrangian, taking into account the relations among the different
sectors and the mixing parameters. We choose to work in an on-shell
renormalization scheme, in which the renormalized parameters are the
measured quantities. 
The SM sector is renormalized according to the                                
standard on-shell SM $\alpha$-scheme~\cite{Hollik}, and the MSSM Higgs         
sector (in particular the renormalization of $\tb$) is treated as in~\cite{Dabels}.

As far as the sfermion sector is concerned, we follow the procedure
described in~\cite{Guasch:1998as}. However,  in the present
analysis we treat simultaneously top-squarks and bottom-squarks. Due to
$SU(2)_L$ invariance the parameters in these two sectors are not
independent, and we can not supply with independent on-shell conditions
for both sectors. We choose as input parameters the on-shell masses of
both bottom-squarks, the lightest top-squark mass, and the mixing angles
in both sectors\footnote{Throughout this work we make use of third generation
notation. The notation is as in~\cite{Guasch:1998as,EWcorr}.}: 
\begin{equation}
  \label{eq:inputsf}
(m_{\tilde{b}_1}, m_{\tilde{b}_2}, \osb ,m_{\tilde{t}_2},
\ost), \ \ \ \ m_{\sfr_1}>m_{\sfr_2}.  
\end{equation}
 The remaining parameters are computed as a function of those
 in~(\ref{eq:inputsf}). In particular, the trilinear soft-SUSY-breaking
 couplings read:
\begin{equation}
A_{\{b,t\}}=\mu\{\tan\beta,\cot\beta\}+
{m_{\tilde{f}_1}^2-m_{\tilde{f}_2}^2\over 2\,m_f}\,\sin{2\,\osf}\,,
\label{eq:Abt}
\end{equation}
with $\tb=v_2/v_1$, the ratio of the vacuum expectation values of the
two Higgs boson doublets. The approximate (necessary) condition to avoid
colour-breaking minima in the MSSM Higgs potential~\cite{Frere:1983ag},
\begin{equation}
A_q^2<3\,(m_{\tilde{t}}^2+m_{\tilde{b}}^2+M_H^2+\mu^2)\,,
\label{eq:necessary}
\end{equation}
imposes a tight correlation between the sfermion mass splitting and the
mixing angle at large $\tb$.
 Since the heaviest top-squark mass ($\msto$)
is not an input 
parameter, it 
receives finite radiative corrections:
\begin{equation}
  \label{eq:mst1radcor}
  \Delta \msto^2 = \delta \msto^2 + \Sigma_{\stopp_1}(\msto^2) \,\,, 
\end{equation}
where $\delta\msto^2$ is a combination of the counterterms of the
parameters in~(\ref{eq:inputsf}), and the counterterms of the gauge and
Higgs sectors. 

The chargino/neutralino sector contains six particles, but only three
independent input parameters:   the soft-SUSY-breaking
$SU(2)_L$ and $U(1)_Y$ gaugino masses ($M$ and $M'$), and the higgsino mass parameter ($\mu$). The situation in this sector is quite
different from the 
sfermion case,
since in this case no independent
counterterms for the mixing matrix elements can be introduced. We stick
to the following procedure: First, we introduce a set of
\textit{renormalized} parameters $(M,M',\mu)$ in the expression of the
chargino and neutralino matrices ($\cal{M}$ and ${\cal M}^0$), and
diagonalize them  by means of unitary matrices $M_D=U^* {\cal
  M} V^\dagger$, $M_D^0 = N^* {\cal M}^0 N^{\dagger}$. Now $U$, $V$ and
$N$ must be regarded as \textit{renormalized mixing matrices}. The
counterterm mass matrices are then $\delta M_D=U^* \delta{\cal M}
V^\dagger$, $\delta M_D^0 = N^*\delta {\cal M}^0 N^{\dagger}$, which are
non-diagonal. At this point, we introduce renormalization conditions for
certain elements of $\delta M_D$ and $\delta M_D^0$. In particular, we use
on-shell renormalization conditions for the two chargino masses ($M_1$
and $M_2$), which allows to compute the counterterms $\delta M$ and
$\delta \mu$. This information, together with the on-shell condition for
the lightest neutralino mass ($M_1^0$) allows to derive the expression
for the counterterm $\delta M'$.
The other
neutralino masses ($M_{2,3,4}^0$) receive radiative corrections. In this
framework the 
renormalized one-loop 
chargino/neutralino 2-point functions 
are \textit{non}-diagonal. Therefore one must take into account this
mixing either by including explicitly the reducible $\chi_r-\chi_s$
mixing diagrams, or by means of  external mixing wave-function terms
(${\cal Z}_{\{L,R\}}^{0\beta \alpha}$, ${\cal Z}_{\{L,R\}}^{-ij}$).
See Refs.~\cite{EberlFritzsche} 
for different (but
one-loop equivalent) approaches to the renormalization of the
chargino/neutralino sector.\footnote{See Ref.~\cite{Majerotto:2002iu} for a
  review of radiative corrections to SUSY processes.}

The complete one-loop computation consists of: 
\begin{itemize}
\item 
renormalization constants for the parameters and wave functions in
  the bare Lagrangian, 
\item
one-loop one-particle irreducible three-point functions, 
\item 
mixing terms among the external charginos and neutralinos, 
\item 
soft- and hard- photon  bremsstrahlung.
\end{itemize}
All kind of MSSM particles are taken into account in the loops: SM fermions,
sfermions, electroweak gauge bosons, Higgs bosons, Goldstone bosons,
Fadeev-Popov ghosts, charginos, neutralinos. The computation is
performed in the 't Hooft-Feynman gauge, using dimensional reduction
for the regularization of divergent integrals. 
The loop computation itself is done using the computer algebra packages
\textit{FeynArts 3.0} and \textit{FormCalc
  2.2}~\cite{FeynArts3,Hahn:1998yk}. 
The numerical evaluation of one-loop integrals makes
use of \textit{LoopTools 1.2}~\cite{Hahn:1998yk}.\footnote{The resulting
  FORTRAN code can be obtained from~\cite{Program}.}

\section{Results}

The results show the very interesting property that none of the particles
of the MSSM decouples from the corrections to the
observables~(\ref{eq:gammadef}). This can be well understood in terms of
renormalization group (RG) running of the parameters and SUSY
breaking. Take, e.g., the effects of squarks in the
electron-selectron-photino coupling. Above the squark mass scale
($Q>m_{\sq}$) the electron electromagnetic coupling ($\alpha(Q)$) is equal
(by SUSY) to the electron-selectron-photino coupling
($\tilde{\alpha}(Q)$), and both couplings run according to the same RG
equations. At $Q=m_{\sq}$ the squarks \textit{decouple} from the
RG running of the couplings. At $Q<m_{\sq}$, $\alpha(Q)$ runs due to the
contributions from pure quark loops, but $\tilde{\alpha}(Q)$ does not
run anymore, and it is
\textit{frozen} at the squark scale, that is:
$\tilde{\alpha}(Q<m_{\sq})=\alpha(m_{\sq})$.  Therefore, when comparing
these two couplings at a scale $Q<m_{\sq}$, they differ by  the
logarithmic running of $\alpha(Q)$ from the squark scale to $Q$:
$\tilde{\alpha}(Q)/\alpha(Q)-1=\beta \log(m_{\sq}/Q)$.

The above discussion has two important consequences: 
\begin{enumerate}
\item The
non-decoupling can be used to extract information of the high-energy
part of the SUSY spectrum: one can envisage a SUSY model in which a
significant splitting among the different SUSY masses exists,
e.g. $m_{\sq} \gg m_{\tilde l}$, where the sleptons lie below the
production threshold in an $e^+e^-$ linear collider, but the squarks are
above it. By means of high precision measurements of the
lepton-slepton-chargino/neutralino couplings one might be able to
extract information of the squark sector of the model, to be checked
with the available data from the LHC. 
\item  By the same token, it means
that the value of the radiative corrections depends on all
parameters of the model, and we can not make precise quantitative
statements unless the full SUSY spectrum is known. 
This drawback can be partially overcome by the introduction of                
\textit{effective coupling                                                     
matrices}, which can be defined as follows. The subset of fermion-sfermion     
one-loop contributions to                                                      
the self-energies of gauge-boson, Higgs-bosons, Goldstone-bosons, charginos    
and neutralinos form a gauge invariant finite subset of the corrections.       
Therefore these contributions can be absorbed into a finite shift of the       
chargino/neutralino mixing matrices $U$, $V$ and $N$
appearing in the
couplings: 
\begin{equation}
  U^{eff}=U+\Delta U^{(f)}   , \ 
  V^{eff}=V+\Delta V^{(f)}    ,  \   N^{eff}=N+\Delta N^{(f)}.
\label{eq:effectivegen}
\end{equation}
In this
  way we can \textit{decouple} the computation of the \textit{universal}
  (or \textit{super-oblique}~\cite{Katz:1998br}) corrections. These
  corrections contain the 
  non-decoupling logarithms from sfermion masses. 
\end{enumerate}
As an example of the \textit{universal} corrections we have computed
the electron-selectron contributions to the $\Delta U^{(f)}$ and $\Delta V^{(f)}$
matrices, assuming zero mixing angle in the
selectron sector ($\theta_e=0$), we have identified the leading terms in the
approximation $m_{\tilde e_i}, m_{\tilde \nu}\gg (\mw,\mi) \gg m_e$, and
 analytically canceled the divergences and the  renormalization scale
 dependent terms;
finally, we have kept only the terms logarithmic in the {slepton}
masses. The result for $\Delta U^{(f)}$ reads as follows:
\begin{eqnarray}
\Delta\UChaio^{(f)}&=& \commonnumfactor\log\left(\frac{M^2_{\tilde e_L}}{\mxs}\right)\,\bigg[
 \frac{\UChaio^3}{6} - 
  \UChait \frac{\sqrt{2}\,\mw\,(M\,\cbta +
  \mu\,\sbta)}{3\,(M^2-\mu^2)\,\mchasomenysmchast^2} 
\left(M^4 - M^2\,\mu^2 + \right.\nonumber\\
&&\left. + 3\,M^2\,\mws + 
     \mu^2\,\mws + \mwf + \mwf\,\cfbt 
    + (\mu^2-M^2)\,\mis + 
     4\,M\,\mu\,\mws\,\stbt\right)\,\bigg]\,\,,
\nonumber\\
\Delta\UChait^{(f)}&=&\commonnumfactor\log\left(\frac{M^2_{\tilde e_L}}{\mxs}\right)\,
\UChaio\,\frac{\mw\,(M\,\cbta + \mu\,\sbta)}
 {3\,\sqrt{2}\,(M^2-\mu^2)\,\mchasomenysmchast^2} \times \nonumber\\
  &\times&  \left((M^2-\mu^2)^2 
+ 4\,M^2\,\mws + 4\,\mu^2\,\mws + 2\,\mwf + 
   2\,\mwf\,\cfbt + 8\,M\,\mu\,\mws\,\stbt\right)\,\,,
\label{eq:logterms}
\end{eqnarray}
$M^2_{\tilde e_L}$ being the soft-SUSY-breaking mass of the
$(\tilde{e}_L,\tilde{\nu})$ doublet,
whereas $\mx$ is a SM mass.
In the on-shell scheme for the SM electroweak theory we define
parameters at very different scales, basically $\mx=\mw$ and
$\mx=m_e$.  These wide-ranging scales enter the structure of the
counterterms and so
must appear in eq.(\ref{eq:logterms}) too. As a result the leading log
in the various terms of this equation will vary accordingly. For
simplicity in the notation we have factorized $\log M^2_{\tilde
e_L}/\mxs$ as an overall factor. In some cases this factor can be very
big, $\log M^2_{\tilde e_L}/m_e^2$; it comes from the electron-selectron
contribution to the chargino-neutralino self-energies.  

\begin{figure}[t]
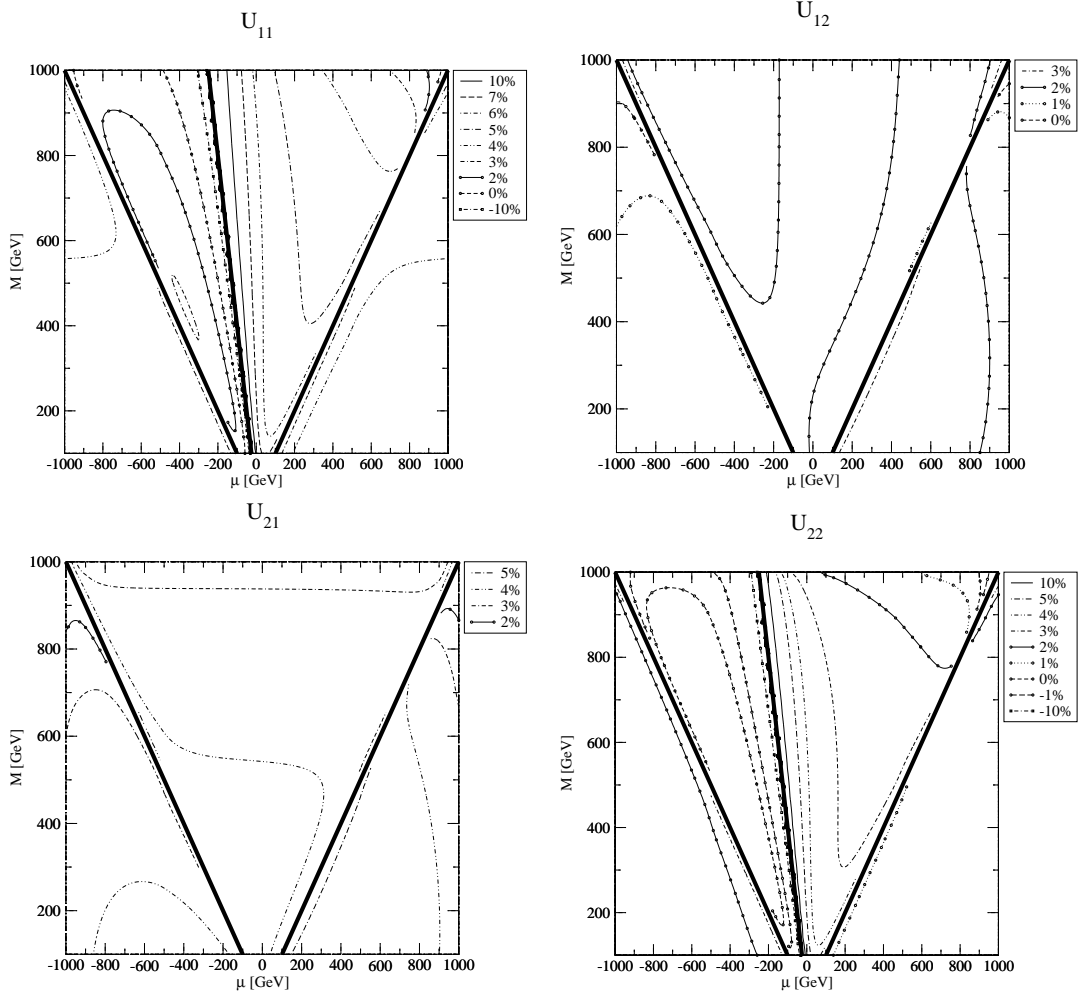

  \centering
\begin{tabular}{cc}
\resizebox{6.9cm}{!}{\includegraphics*{DUeff11.eps}}
&
\resizebox{6.9cm}{!}{\includegraphics*{DUeff12.eps}}
\\
\resizebox{6.9cm}{!}{\includegraphics*{DUeff21.eps}}
&
\resizebox{6.9cm}{!}{\includegraphics*{DUeff22.eps}}
\end{tabular}
  \caption{Relative correction to the effective chargino coupling matrix $\Delta U^{(f)}/U$
     in the $M-\mu$ plane, for $\tb=4$ and a
    sfermion spectrum around $1\TeV$~($m_{\tilde{l}_2}=m_{\tilde{d}_2}=m_{\tilde{u}_2}=1 \TeV \,,\,
 m_{\tilde{l}_1}=m_{\tilde{d}_1}=m_{\tilde{l}_2}+5\GeV\,,\,
\theta_l=\theta_q=\theta_b=0 \,,\, \theta_t=-\pi/5$).} 
  \label{fig:Ueff}
\end{figure}

In Fig.~\ref{fig:Ueff} we show the relative correction to the matrix
elements of $U$ for a sfermion spectrum around $1\TeV$. 
The thick black lines in Fig.~\ref{fig:Ueff} correspond
to {spurious} divergences in the relative corrections due to the
renormalization prescriptions.
Corrections as
large as $\pm10\%$ can only be found in the 
vicinity of these divergence lines. However, there exist large regions
of the $\mu-M$ plane  where the corrections are larger than $2\%$,
$3\%$, or even $4\%$.

\begin{figure}[tp]
  \begin{center}
    \begin{tabular}{cc}
\resizebox{7cm}{!}{\includegraphics{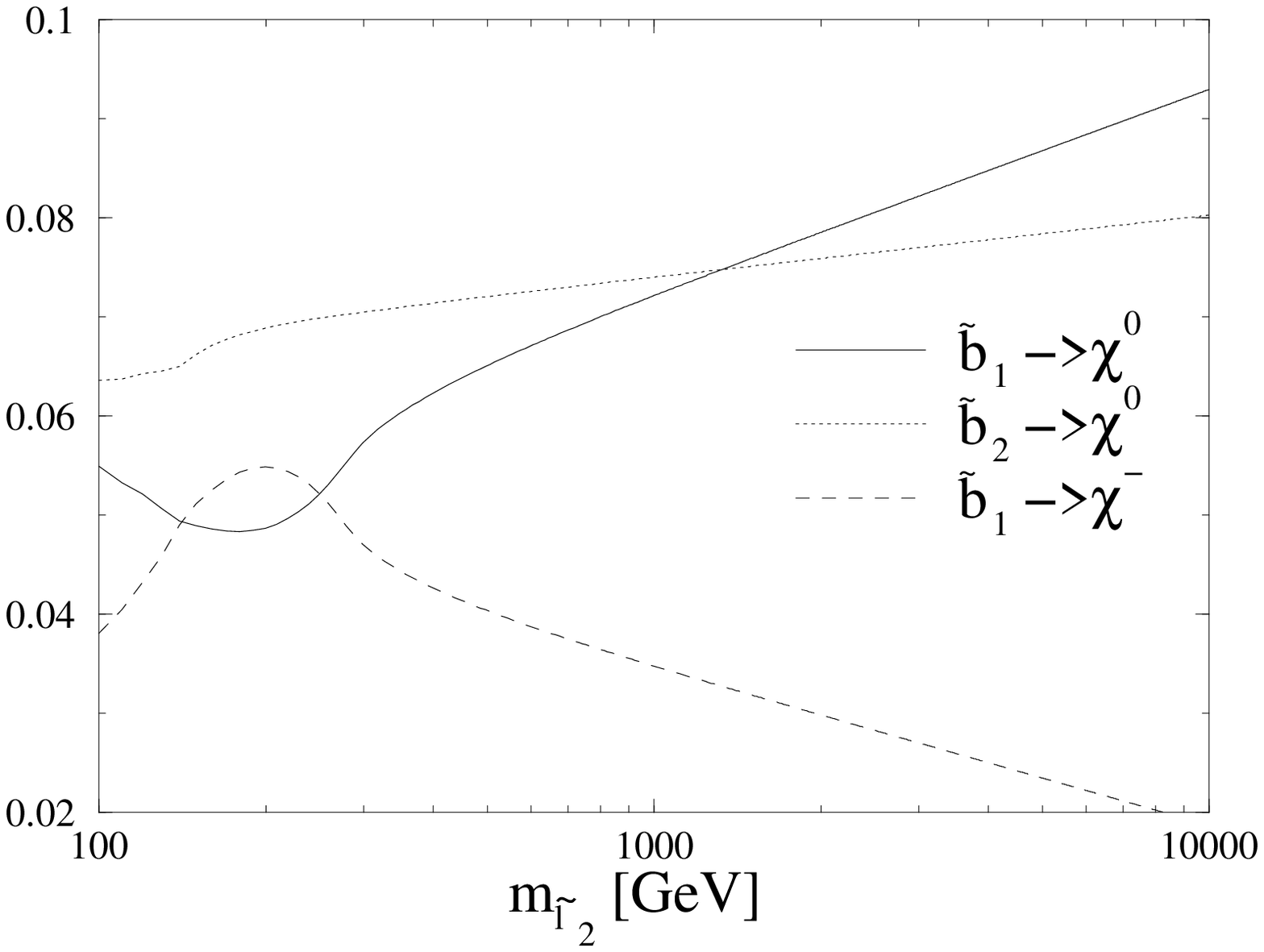}}&
\resizebox{7cm}{!}{\includegraphics{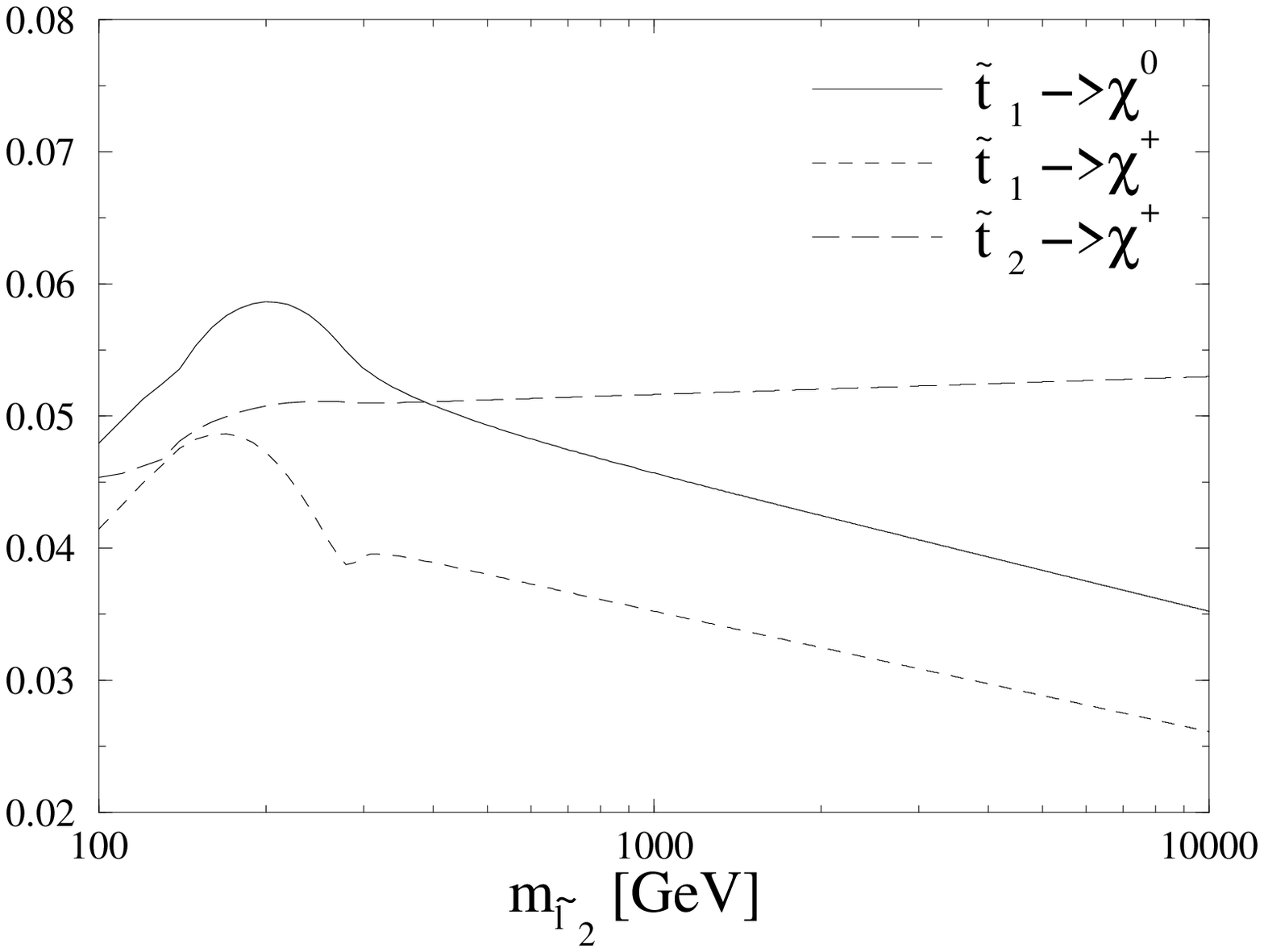}}
\\(a)&(b)
    \end{tabular}
  \end{center}
\caption{Universal relative corrections~(\ref{eq:totalcn}) to third generation squark partial decay widths as a
  function of a common slepton mass using the input parameter set~(\ref{eq:inputpars}).}  \label{fig:unisq}
\end{figure}

The effects of the universal corrections to the partial decay
widths~(\ref{eq:gammadef}) are shown in Fig.~\ref{fig:unisq} for top-
and bottom-squark decays as a function of a common slepton mass. 
Here (and in most of the 
discussion below) we show
the corrections to the total decay widths of sfermions into charginos and
neutralinos, that is
\begin{equation}
  \label{eq:totalcn}
\delta(\tilde{f_a}\to f'\chi)=\frac{ \sum_r 
    \left(\Gamma(\tilde{f_a}\to  f'\chi_r)-\Gamma^0(\tilde{f_a}\to  f'\chi_r)\right)}
  {\sum_r \Gamma^0(\tilde{f_a}\to  f'\chi_r)}\,\,, 
\end{equation}
with $\chi=\chi^\pm$ or $\chi=\neut$. 
We will not show results for processes {whose} branching ratio
{are} less that 10\% in all of the explored parameter space. The default
parameter set used is:
\begin{equation}
\begin{array}{l}
 \tb      =  4\,\,,\mt=175\GeV\,\,,
 \mb=5\GeV\,\,,
 \msbt=\msdt=\msut=\mselt=300\GeV\,\,,\\
 \msbo=\msdo=\mselo=\msbt+5\GeV\,\,,
\msut=290\GeV\,\,,
 \mstt=300\GeV\,\,,\\
 \osb=\osd=\osu=\osel=0\,\,,
 \ost=-\pi/5\,\,, 
 \mu      =  150\GeV\,\,,
 M       =  250\GeV\,\,,
 \mHp  =  120\GeV\,\,,
\end{array}
\label{eq:inputpars}
\end{equation}
The logarithmic behaviour from eq.~(\ref{eq:logterms}) is evident in this figure. 
{The logarithmic regime is attained already for slepton masses of
  order $1\TeV$.}
The universal corrections are seen to be
positive for all squark decays, ranging between $4\%$ and $7\%$ for
slepton masses below $1\TeV$.

Although above we have singled out the non-decoupling properties of
sfermions, we would like to stress that the whole spectrum shows
non-decoupling properties. By numerical analysis we have been able to
show the existence of logarithms of the gaugino mass parameters ($M/M_X$ and
$M'/M_X$), and the Higgs mass ($\mHp/M_X$). However, due to the
complicated mixing 
structure of the model, we were not able to derive simple analytic
expressions containing these non-decoupling logarithms. 
Note that in \textit{any} observable which includes the
fermion-sfermion-chargino/neutralino Yukawa couplings at leading order
we will have this 
kind of corrections, therefore the full MSSM spectrum must be taken into
account when computing radiative corrections, since otherwise one could
be missing large logarithmic contributions of the heavy masses.

\begin{figure}[tbp]
  \begin{center}
    \begin{tabular}{cc}
\resizebox{7cm}{!}{\includegraphics{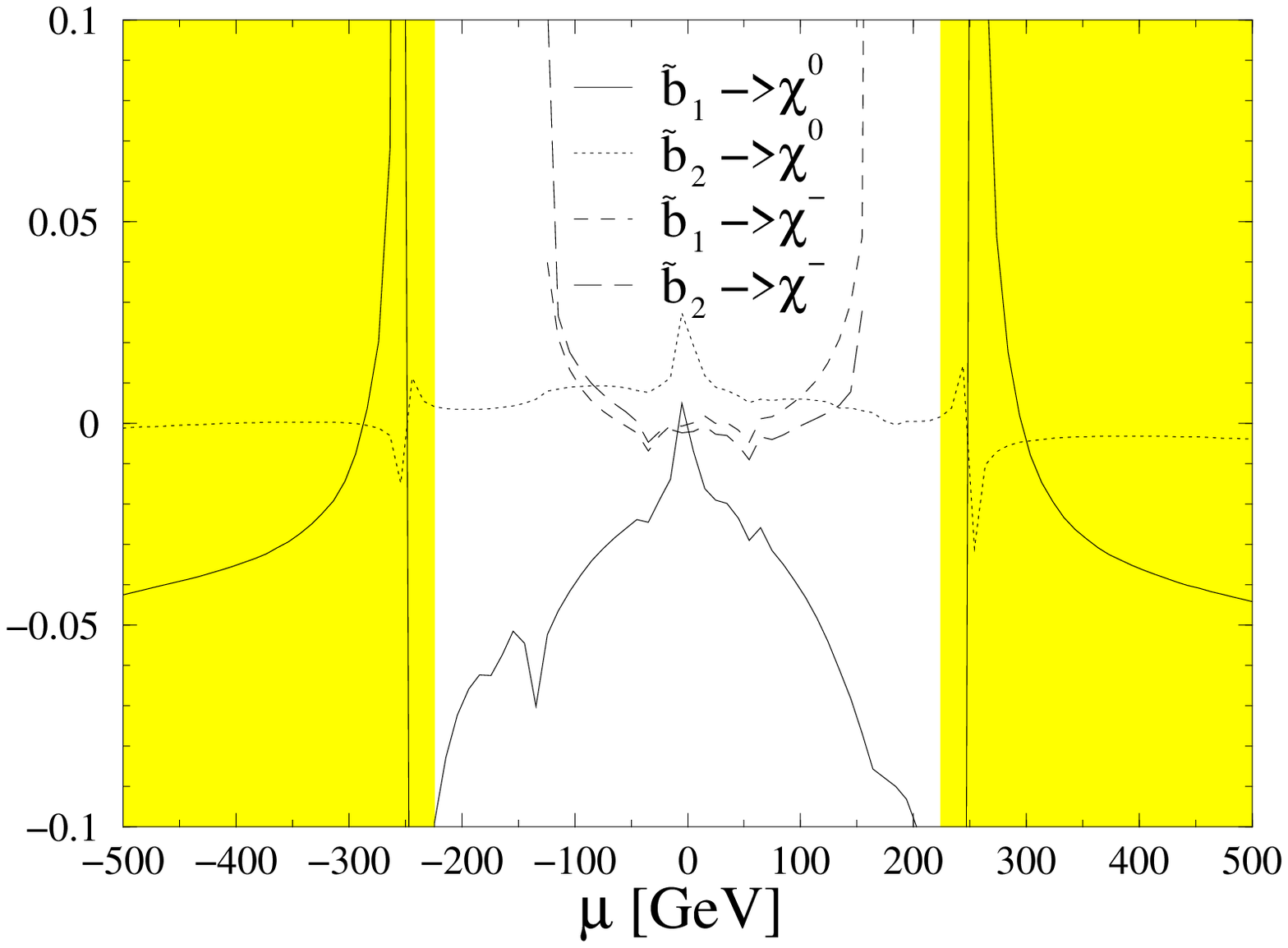}}&
\resizebox{7cm}{!}{\includegraphics{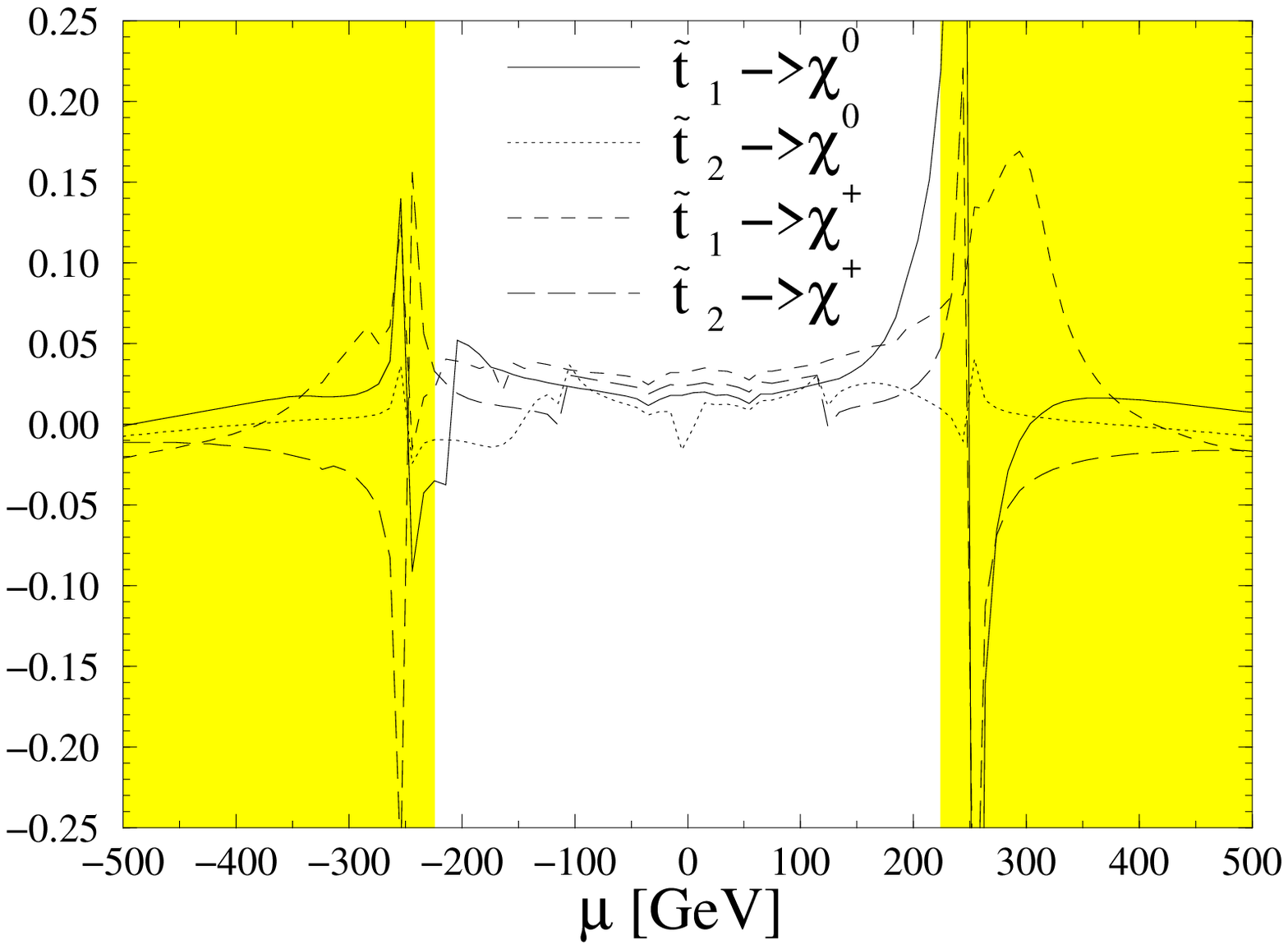}}
\\(a)&(b)
    \end{tabular}
  \end{center}
  \caption{Non-universal corrections to the partial decay width of
top- and bottom-squarks as a function of the higgsino mass parameter
    $\mu$. 
  {The shaded regions correspond to the violation of the
  condition~(\ref{eq:necessary}).}}\label{fig:muquark}
\end{figure}

As for the \textit{non-universal} part of the contributions, they show a
rich structure, as can be seen in Fig.~\ref{fig:muquark}. There we show
the evolution of the corrections as a function of the $\mu$ parameter
for top- and bottom-squark decays. 
A number of divergences are seen in the figure,
ones related to the mass renormalization framework (at $|\mu|=M$), and others
due to threshold singularities in the external wave function
renormalization constants. It is clear that the precise value of the
corrections is very much dependent on the correlation among the
different SUSY masses.

An important contribution to the corrections of third-generation
sfermion decays is the \textit{threshold correction} to the bottom-quark
($\tau$-lepton) Yukawa coupling 
($\Delta m_{\{b,\tau\}}$)~\cite{DmbTeo}. In the processes under
study~(\ref{eq:gammadef}) two kind of contributions appear: first, the
genuine corrections $\Delta m_{\{b,\tau\}}$ from SUSY loops in the
fermion self-energy; and second in the loops of sfermion self-energies
mixing different chiral states $\sfr_L \leftrightarrow \sfr_R$. This
kind of corrections grow with the sfermion mass splitting, the sfermion mixing
angle, and $\tb$. 

\begin{figure}
  \centering
\begin{tabular}{cc}
\resizebox{7cm}{!}{\includegraphics{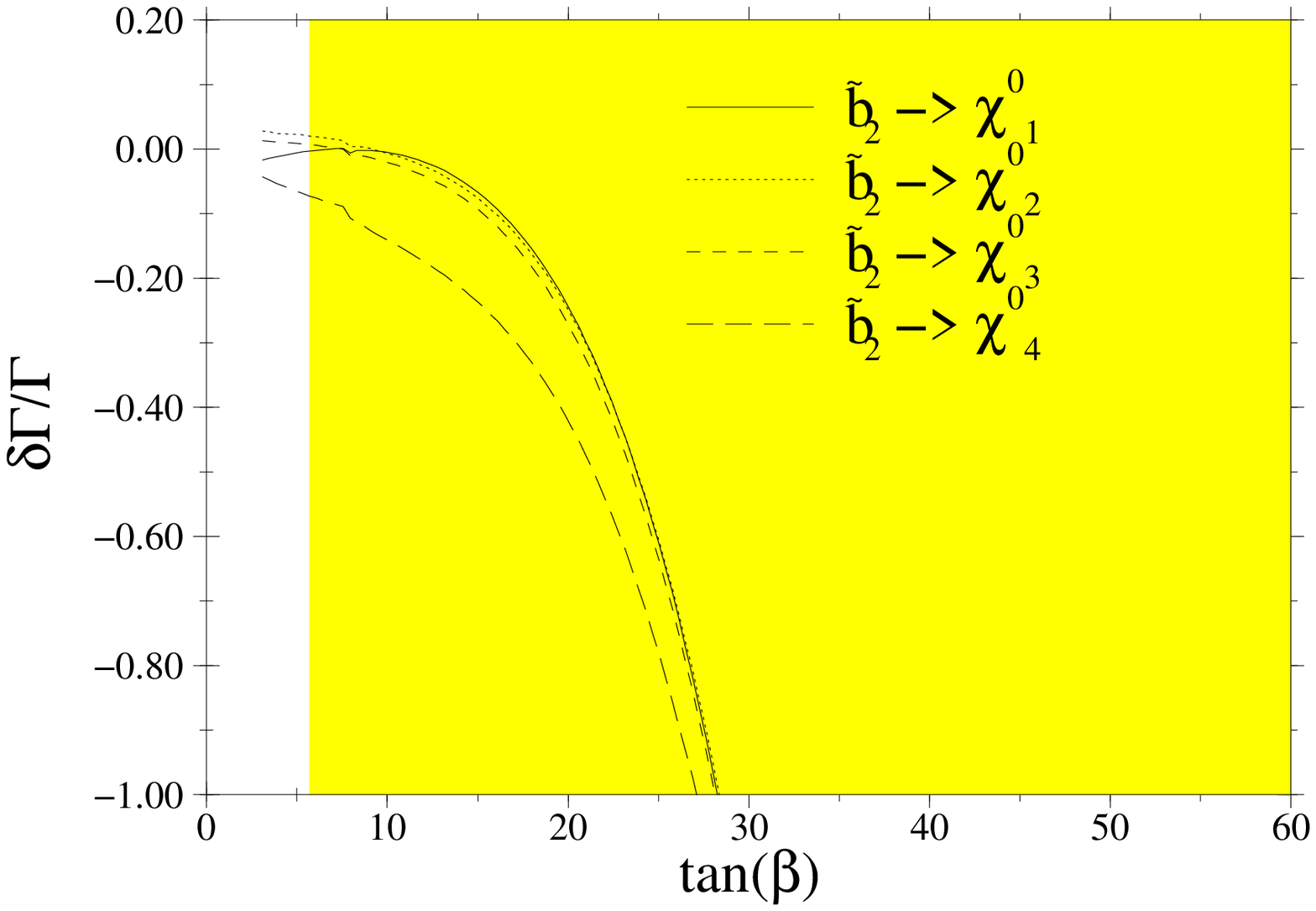}} &
\resizebox{7cm}{!}{\includegraphics{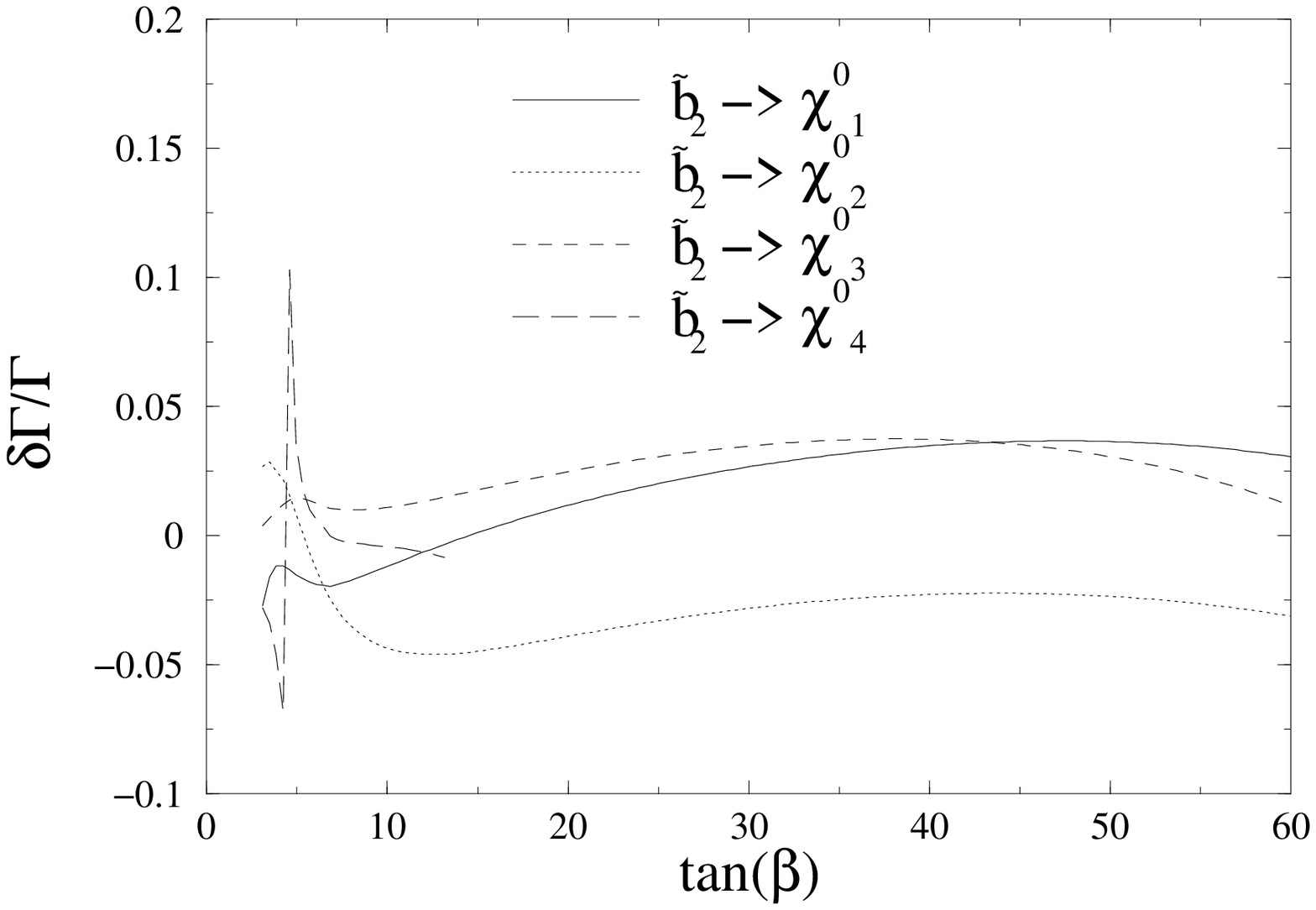}} \\
(a) & (b) 
\end{tabular}  
\caption{Non-universal relative corrections to the lightest
  bottom-squark partial decays 
  widths into  neutralinos as a function of
  $\tb$. \textbf{a)} Keeping fixed the splitting between the
  bottom-squarks $\msbo-\msbt=5\GeV$. \textbf{b)} Keeping $A_b=600\GeV,
  A_t=-78\GeV$.   {The shaded region corresponds to the violation of the
  condition~(\ref{eq:necessary}).}}  
\label{fig:sbottomcomp}
\end{figure}

A complementary set of corrections corresponds to the genuine
three-point vertex functions including Higgs bosons in the loops. These
contributions are proportional to the soft SUSY-breaking trilinear
couplings~(\ref{eq:Abt}), and therefore potentially large. Concretely,
if $\tb$ is large, and the bottom-squark mass splitting (or the mixing
angle) is small, the bottom-squark trilinear coupling grows with $\tb$
($A_b\simeq \mu\tb$), eventually inducing corrections larger than
$100\%$, spoiling the validity of perturbation theory. 
In Fig.~\ref{fig:sbottomcomp}a we show the evolution of the
corrections to the lightest bottom-squark decay into neutralinos as a
function of $\tb$ using the parameter 
set~(\ref{eq:inputpars}). We see the fast growing of the corrections,
reaching $-100\%$ at $\tb\simeq30$. Fortunately, applying the (necessary) 
restriction~(\ref{eq:necessary}) keeps the $A_q$ parameter small. In
Fig.~\ref{fig:sbottomcomp}b we show again the evolution of the
corrections as a function of $\tb$, but this time keeping a fixed value
for the trilinear couplings $A_b=600\GeV$, $A_t=-78\GeV$. The figure
shows that the corrections stay well below $10\%$ all over the $\tb$
range for this channel.

The complementarity between the $\Delta m_{\{b,\tau\}}$-like and the
$A_f$-like corrections is as follows: at large $\tb$, if the
bottom-squark mass splitting is large, there will be large corrections of
type $\Delta m_{\{b,\tau\}}$; on the other hand, if the bottom-squark
mass splitting is small, there will be large corrections of the type
$A_f$. Note that the QCD corrections contain $\Delta m_b$ terms but not
$A_f$ terms. When analyzing QCD corrections alone, one could choose a
small splitting, obtaining small corrections, however we have seen that
this is inconsistent, so one is forced to a large  $\Delta\mb^{QCD}$
contribution, which can reinforce (or screen) the negative corrections
from the standard running of the QCD coupling constant\footnote{Though
  it is not possible to separate between standard gluon corrections and
  gluino corrections, one can talk qualitatively about the
  contributions of the different sectors.}.

It is known that the electroweak corrections to any process grow as the
logarithm squared of the process energy scale due to the Sudakov
double-logs~\cite{sudakov}. We have observed this behaviour in the
process under study. 

At the end of the day, we want to analyze the branching ratios, which
are the true observables. For this analysis we have to add the QCD
corrections to the EW corrections. Due to the large value of the QCD
corrections, we made use of the enhanced resummed expression for
the bottom-quark Yukawa coupling~\cite{davidDmb}.
In Table~\ref{tab:brcorr} we show the tree-level and corrected branching
ratios for 
top- and bottom-squarks using the input parameter
set~(\ref{eq:inputpars}) and $\mg=500\GeV$. From inspection of
Table~\ref{tab:brcorr} we 
 see that the EW corrections can induce
 a change on the branching ratios of the leading decay channels of squarks
 comparable to the QCD corrections. Therefore both contributions must be
 taken into account on equal footing in the analysis of the phenomenology
 of sfermions.
 \begin{table}[tbp]
 \centering
   \begin{tabular}{|c|c|c|c|c||c|c|}
 \hline  
  & $\neut_1$ & $\neut_2$ & $\neut_3$ & $\neut_4$ & $\cplus_1$ & $\cplus_2$\\ \hline  
 $BR^{tree}(\tilde{t}_1 \to q\chi)$  & 0.169 & 0.249 & 0.145 & - & 0.159 & 0.278\\ \hline  
 $BR^{QCD}(\tilde{t}_1\to q\chi)$ & 0.164 & 0.257 & 0.144 & - & 0.099 & 0.335\\ \hline 
 $BR^{total}(\tilde{t}_1\to q\chi)$ & 0.177 & 0.242 & 0.143 & - & 0.122 & 0.316\\ \hline \hline 
 $BR^{tree}(\tilde{t}_2 \to q\chi)$ & 0.058 & - & - & - & 0.942 & -\\ \hline  
 $BR^{QCD}(\tilde{t}_2\to q\chi)$ & 0.063 & - & - & - & 0.937 & -\\ \hline 
 $BR^{total}(\tilde{t}_2\to q\chi)$ & 0.065 & - & - & - & 0.935 & -\\ \hline \hline 
 $BR^{tree}(\tilde{b}_1 \to q\chi)$  & 0.272 & 0.092 & 0.047 & 0.014 & 0.575 & -\\ \hline  
 $BR^{QCD}(\tilde{b}_1\to q\chi)$ & 0.308 & 0.104 & 0.031 & 0.018 & 0.538 & -\\ \hline 
 $BR^{total}(\tilde{b}_1\to q\chi)$ & 0.291 & 0.092 & 0.031 & 0.018 & 0.568 & -\\ \hline \hline 
 $BR^{tree}(\tilde{b}_2 \to q\chi)$ & 0.502 & 0.332 & 0.123 & - & 0.042 & -\\ \hline  
 $BR^{QCD}(\tilde{b}_2\to q\chi)$ & 0.541 & 0.386 & 0.054 & - & 0.019 & -\\ \hline 
 $BR^{total}(\tilde{b}_2\to q\chi)$ & 0.528 & 0.395 & 0.056 & - & 0.020 & -\\ \hline 
 \end{tabular}
 \caption{Tree-level and corrected branching ratios of top- and
 bottom-squark decays 
   into charginos and 
  neutralinos for the parameter set~(\ref{eq:inputpars}) and
   $\mg=500\GeV$.  Branching 
  ratios below $10^{-3}$ are not shown.} 
\label{tab:brcorr}
\end{table}

\section{Squark effects in sleptons observables}

Since the corrections do not decouple by taking the large mass limit,
the logical question appears: which knowledge of the heavy spectrum is
necessary in order to provide a theoretical prediction with sufficient
accuracy for the properties of the light particles? In the following we
try to answer this question. To this end, we choose a spectrum with
light sleptons and heavy squarks, and look at the radiative effects of
the latter in the properties of the former. For the numerical analysis we
choose as default parameters those of the \textit{Snowmass Points and
  Slopes} (SPS), point 1a~\cite{Allanach:2002nj}\footnote{The spectrum and tree-level branching
  ratios for the several SPSs can be found e.g. in
Ref.~\cite{Ghodbane:2002kg}.}. For completeness we give here the values of the
soft-SUSY-breaking parameters for this point:
\begin{equation}
\begin{array}{l} 
\tb= 10, 
 \mA= 393.6 \GeV,  \mu= 352.4\GeV,
 M= 192.7\GeV,
 M'=99.1\GeV\,\,,
\\
 M_{\{\tilde{d},\tilde{s}\}_L}= 539.9 \GeV,
 M_{\tilde{b}_L}=495.9\GeV,
  M_{\{\tilde{d},\tilde{s}\}_R}=519.5\GeV,
  M_{\tilde{b}_R}=516.9\GeV,\\
 A_{\{d,s\}}= 3524\GeV,
 A_b=-772.7\GeV,\\
 M_{\{\tilde{u},\tilde{c}\}_R}= 521.7\GeV,
 M_{\tilde{t}_R}= 424.8\GeV,
 A_{u,c}= 35.24 \GeV,
 A_t=-510\GeV, \\
 M_{\{\tilde{e},\tilde{\mu}\}_L} =196.6\GeV,
 M_{\tilde{\tau}_L}=195.8\GeV,
 M_{\{\tilde{e},\tilde{\mu}\}_R} =136.2\GeV,
 M_{\tilde{\tau}_R}=133.6\GeV,\\
 A_{e,\mu}= 3524\GeV,
 A_\tau=-254.2\GeV,
\end{array}
\label{eq:sps1a}
\end{equation}
where the soft-SUSY-breaking trilinear couplings of the first and second
generation sfermions have been chosen such that the non-diagonal
elements of the sfermion mass matrix are zero.

However, a note of caution should be given, our
computation is performed in the On-shell renormalization scheme, whereas
the SPS parameters are given in the $\overline{DR}$ renormalization
scheme, and one should make a scheme conversion of the
parameters, this conversion is beyond the scope of the present work. In
this note we are interested only in establishing whether the effects of heavy
particles are important, and therefore we are only interested in
obtaining a suitable SUSY spectrum, therefore we treat the given
numerical parameters of SPS 1a as On-shell SUSY 
parameters~\footnote{Of course, once we will be analyzing the real LC
  data, the $\overline{DR}$-On-shell conversion will need to be made in order
to extract the fundamental soft-SUSY-breaking parameters.}.

\begin{table}[tb]
\begin{center}
\begin{tabular}[c]{|c|c|c|c|c|c|c|c|}
\hline
&$\Gamma^{tree}$ [GeV] & $\delta\Gamma^{(q)}/\Gamma$ & $\delta\Gamma^{(l)}/\Gamma$
& $\delta\Gamma^{no-uni}/\Gamma$ & $\delta\Gamma/\Gamma$ \\ \hline
$\tilde{e}_1\to e^-\chi^0_1$ & 0.110 & 0.043 & 0.032 & 
-0.002 & 0.073\\ \hline
$\tilde{e}_1\to e^-\chi^0_2$ & 0.047 & 0.030 & 0.034 & 
-0.012 & 0.051\\ \hline
$\tilde{e}_1\to \nu_e\chi^-_1$ & 0.081 & 0.026 & 0.033 &
0.006 & 0.065\\ \hline
$\tilde{e}_2\to e^-\chi^0_1$ & 0.194 & 0.052 & 0.034 & 
0.000 & 0.086\\ \hline
$\tilde{\nu}_e\to \nu_e\chi^0_1$ & 0.140 & 0.059 & 0.035 & 
-0.005 & 0.089\\ \hline
$\tilde{\nu}_e\to \nu_e\chi^0_2$ & 0.006 & 0.018 & 0.033 & 
-0.014 & 0.036\\ \hline
$\tilde{\nu}_e\to e^-\chi^+_1$ & 0.016 & 0.024 & 0.033 & 
0.002 & 0.059\\ \hline
\end{tabular}
\end{center}
\caption{Tree-level partial decay widths and relative corrections for
  the selectron and sneutrino decays into charginos and neutralinos for
  SPS 1a.\label{tab:gammas}}
\end{table}

In the following we separate among three different kinds of contributions:
$\delta\Gamma^{(q)}$ are the corrections induced by the quark-squark loops
to the universal corrections in~(\ref{eq:effectivegen}), and are the main
subject of study in this section;
$\delta\Gamma^{(l)}$ are the corrections induced by the lepton-slepton loops
to the universal corrections in~(\ref{eq:effectivegen});
$\delta\Gamma^{no-uni}$ are the non-universal corrections as before.

In Table~\ref{tab:gammas} we show the partial decay widths of selectrons
into charginos/neutralinos for SPS 1a. We show: the tree-level partial
widths $\Gamma^{tree}$; 
the relative corrections induced by quarks-squarks $\delta\Gamma^{(q)}/\Gamma$; 
the relative corrections induced by the lepton-slepton universal
contributions $\delta\Gamma^{(l)}/\Gamma$; 
the process-dependent non-universal contributions
$\delta\Gamma^{no-uni}/\Gamma$; and the total corrections
$\delta\Gamma/\Gamma$.

The universal corrections $\delta\Gamma^{(q)}/\Gamma$ and
$\delta\Gamma^{(l)}/\Gamma$ in Table~\ref{tab:gammas} represent a
correction that will be present whenever a
fermion-sfermion-chargino/neutralino coupling enters a given
observable. The correction $\delta\Gamma^{no-uni}/\Gamma$ represents the
process-dependent part. For the presented observables the non-universal
corrections turn out to be quite small, but this is not necessarily always
the case. From the values of Table~\ref{tab:gammas} it is clear that the
corrections of the quark-squark sector are as large as the corrections
from the (light) lepton-slepton sector, for the presented observables they
amount to a $2-6\%$ relative correction, depending on the particular decay
channel. 

\begin{figure}[tbp]
\centerline{\resizebox{7cm}{!}{\includegraphics{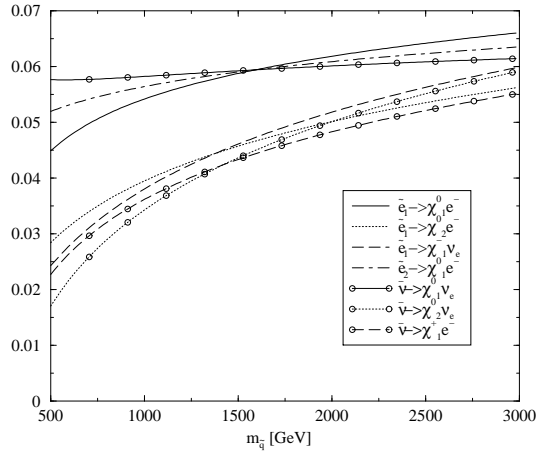}}}
\caption{Squark contributions to the radiative corrections of the
  partial decay widths of sleptons ($\delta\Gamma^{(q)}/\Gamma$) for SPS
  1a as a function of a common soft-SUSY-breaking mass parameter for all
  squarks.\label{fig:UniQmsquark}}
\end{figure}

For SPS 1a the squark mass scale is around 500\GeV, however the
corrections grow logarithmically with the squark mass scale. In
Fig.~\ref{fig:UniQmsquark} we show the relative corrections induced by the
quark-squark sector ($\delta\Gamma^{(q)}/\Gamma$) in the observables of
Table~\ref{tab:gammas} as a function of a common value for all
soft-SUSY-breaking squark mass parameters in~(\ref{eq:sps1a}), in a range where the squarks
are accessible at the LHC. The several lines in
Fig.~\ref{fig:UniQmsquark} are neatly grouped together: the upper lines
correspond to the lightest neutralino ($\neut_1$) which is
\textit{bino}-like, whereas the lower lines correspond to the second
neutralino and lightest chargino ($\neut_2$, $\chi^\pm_1$), which are
\textit{wino}-like. Since the coefficient of the logarithm in the
universal corrections~(\ref{eq:effectivegen}) is
proportional to the corresponding gauge coupling, the behaviour of the
corrections is different between the two kinds of gauginos, but similar
for different gauginos of the same kind. We see that for a
\textit{bino}-like neutralino the corrections undergo an absolute shift
of less than 2\% (from 4.5\% to 6.5\% in the channel
$\tilde{e}_1\to e^-\neut_1$) by changing the squark mass scale from
$500\GeV$ 
to 3\TeV. For a \textit{wino}-like gaugino the shift is much larger,
being up to 4\% in the case under study (from 2\% to 6\% in the
$\tilde{\nu}_e\to\nu_e\neut_2$ channel). We conclude, therefore, that a
certain knowledge of the squark masses is necessary in order to provide
a theoretical prediction with an uncertainty below 1\%, but only a
rough knowledge of the scale is necessary.

\begin{figure}[t]
\begin{tabular}{cc}
\resizebox{6.5cm}{!}{\includegraphics{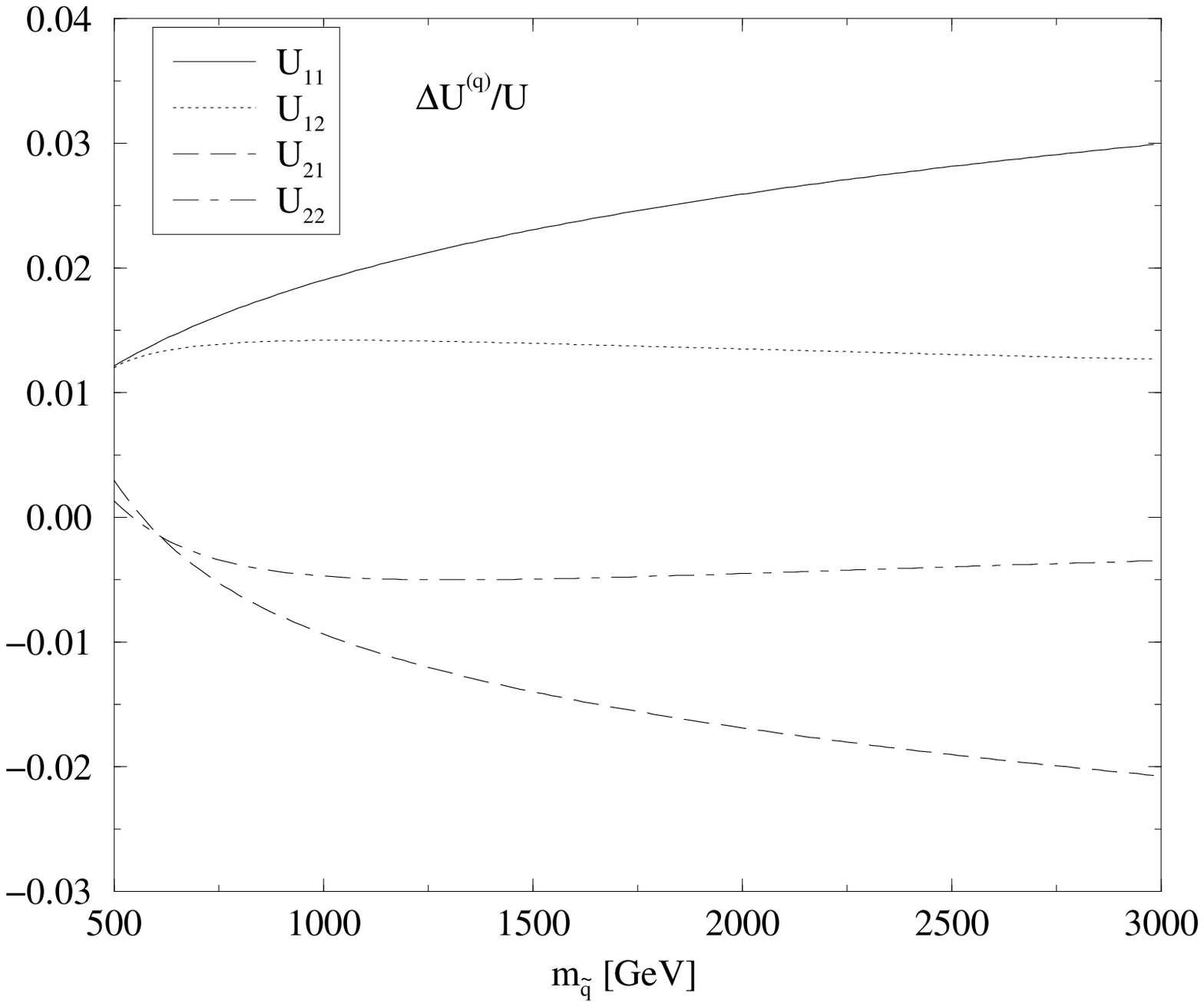}} &
\resizebox{6.5cm}{!}{\includegraphics{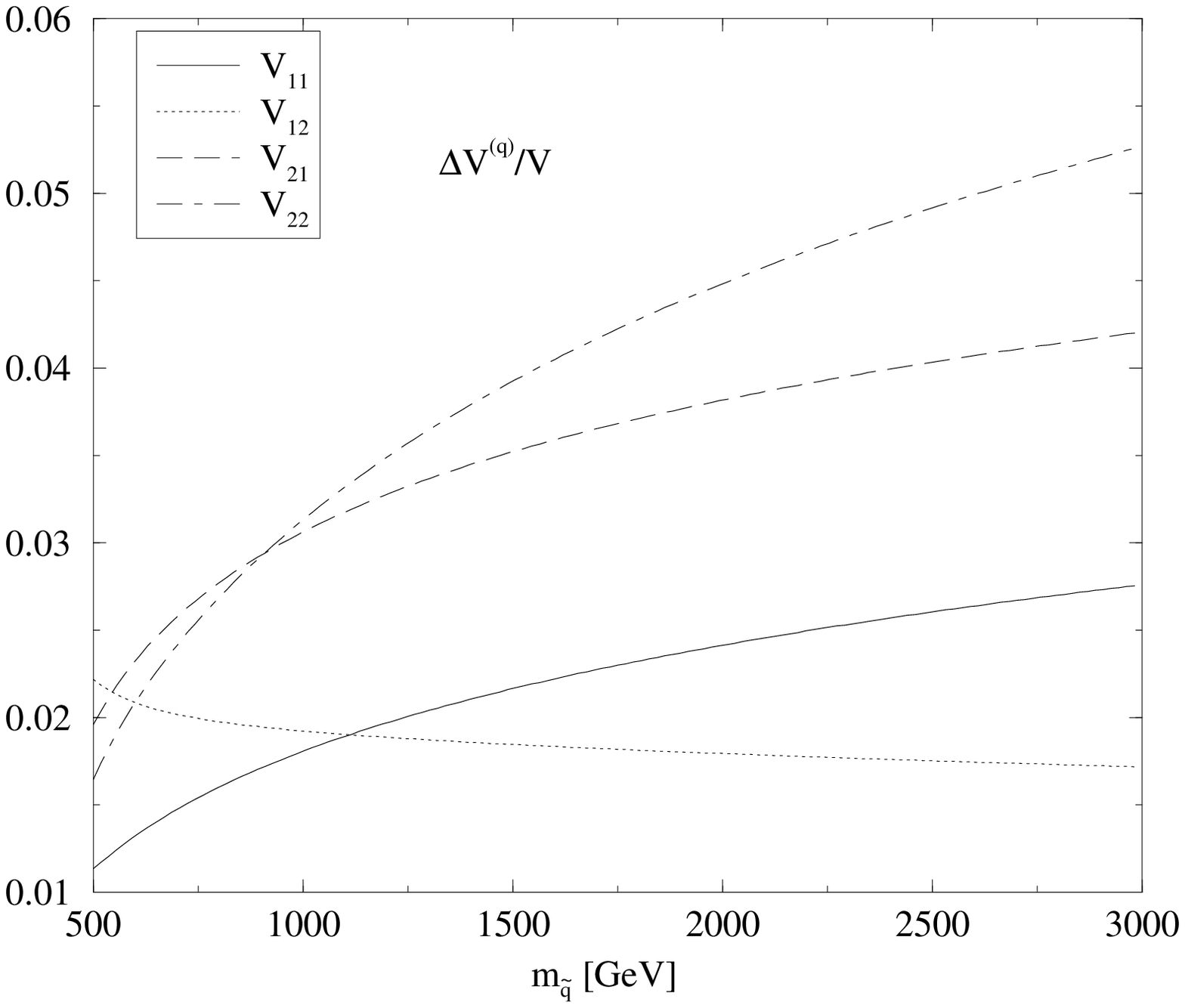}} \\
(a) & (b) \\
\resizebox{6.5cm}{!}{\includegraphics{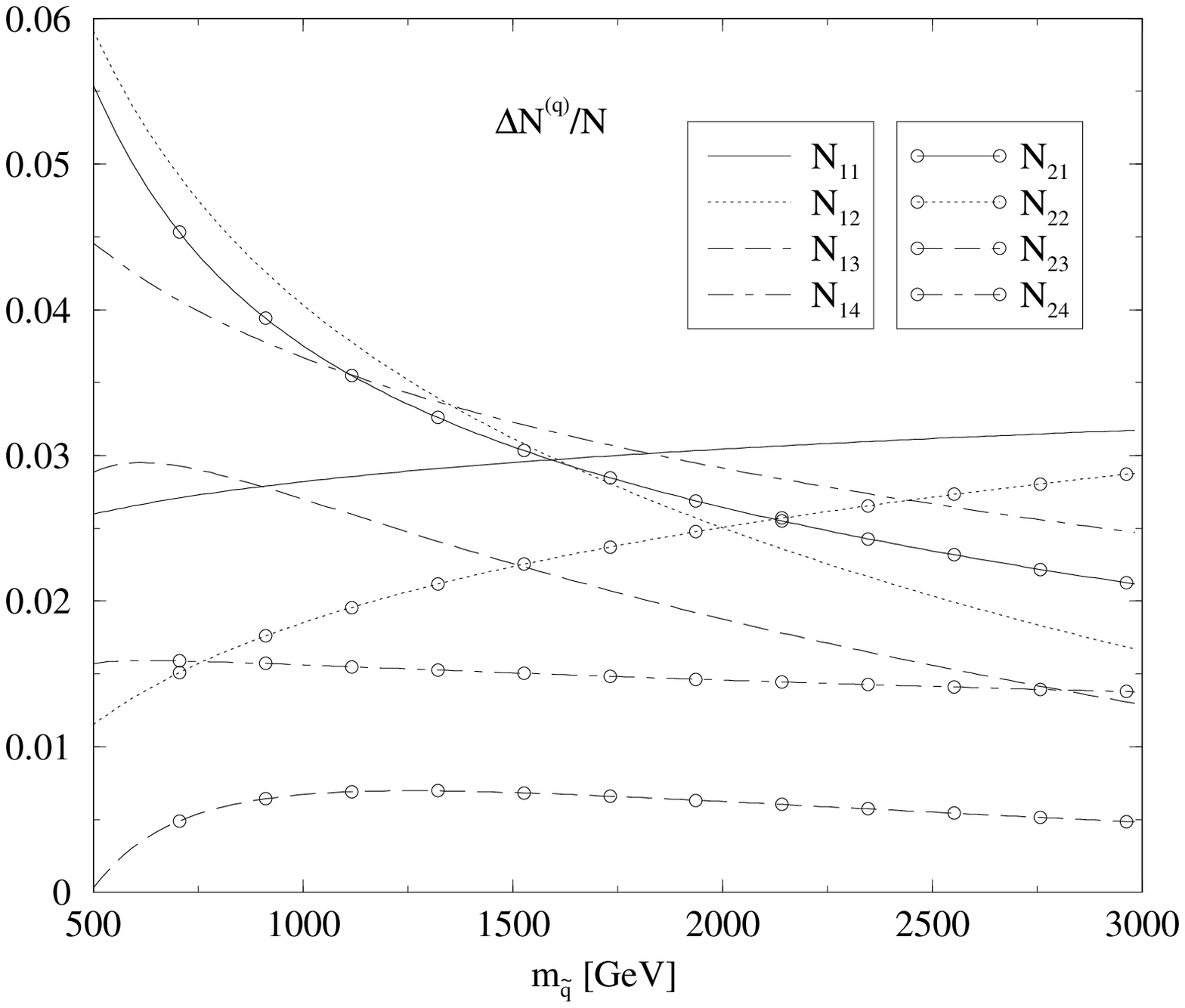}}  &
\resizebox{6.5cm}{!}{\includegraphics{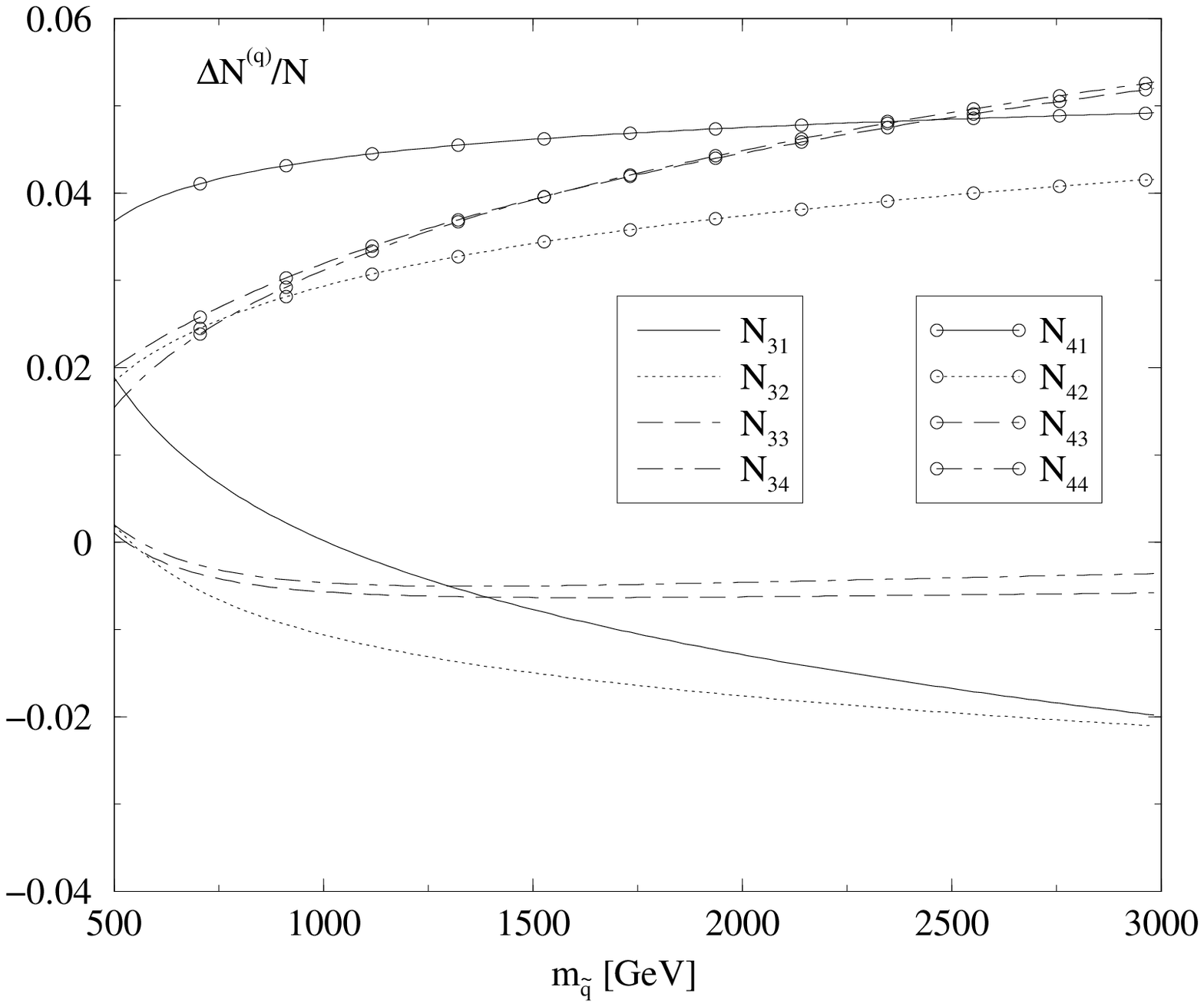}} \\
(c) & (d) 
\end{tabular}
\caption{Quark-squark contributions to the effective chargino/neutralino
  mixing coupling matrices~(\ref{eq:effectivegen}) as a function of a
  common squark mass parameter for SPS 1a.\label{fig:EffQmsq}}
\end{figure}

As explained previously these corrections admit a description in terms
of effective coupling matrices. In Fig.~\ref{fig:EffQmsq} we show the
relative finite shifts induced by the quark-squark sector in the
effective coupling matrices as a function of a common soft-SUSY-breaking
squark mass parameter. The tree-level values for the mixing matrices are:
\begin{equation}
  \label{eq:treeUVN}
\begin{array}{l}
U=\pmatrix{   0.91 &     0.41 \cr
            -0.41 &    0.91 } \, , \, \\
V=\pmatrix{ 0.97 &      0.24 \cr
  -0.24&   0.97 }  \, , \, 
\end{array}
N=\pmatrix{ -0.99        &
  -0.10      &
     -0.06 i &
  0.11      \cr
     0.06 &
     -0.94    &
          0.09 i&
      -0.32           \cr
      -0.15       &
      0.28        &
      0.69 i&
      -0.64      \cr
      0.05    &
      -0.16      &
      0.71 i&
      0.68   }\,\,.
\end{equation}
One can perform a one-to-one matching of Fig.~\ref{fig:EffQmsq} with
Fig.~\ref{fig:UniQmsquark}. By neglecting the small electron-higgsino
couplings we obtain:
\begin{equation}
\begin{array}{rcl}
\delta\Gamma^{(q)}(\tilde{e}_L\to e^-\neut_\alpha)/\Gamma&=&2 \frac{\Delta
N^{(q)}_{\alpha 2} - Y_L t_w \Delta N^{(q)}_{\alpha 1} }
{N_{\alpha 2} - Y_L t_w  N_{\alpha 1} },\nonumber\\
\delta\Gamma^{(q)}(\tilde{e}_L\to\nu_e\cmin_1)/\Gamma&=&2 \Delta U^{(q)}_{11}/U_{11}, \nonumber\\
\delta\Gamma^{(q)}(\tilde{e}_R\to e^-\neut_\alpha)/\Gamma&=&2 \Delta
N^{(q)}_{\alpha 1}/N_{\alpha 1} ,\nonumber\\
\end{array}
\begin{array}{rcl}
\delta\Gamma^{(q)}(\tilde{\nu}_e\to e^-\cplus_1)/\Gamma&=&2 \Delta V^{(q)}_{11}/V_{11}, \nonumber\\
\delta\Gamma^{(q)}(\tilde{\nu}_e\to \nu_e\neut_\alpha)/\Gamma&=&2  \frac{\Delta
N^{(q)}_{\alpha 2} + Y_L t_w \Delta N^{(q)}_{\alpha 1} }
{N_{\alpha 2} + Y_L t_w  N_{\alpha 1} } ;
\end{array}
\end{equation}
and $\tilde{e}_{\{L,R\}}=\tilde e_{\{1,2\}}$ for the
case~(\ref{eq:sps1a}) under 
study. We see in Fig.~\ref{fig:EffQmsq} variations up
to 5\% in the coupling matrices, which would translate to variations
up to 10\% in the observables.

\begin{figure}[tbp]
\begin{tabular}{cc}
\resizebox{7cm}{!}{\includegraphics{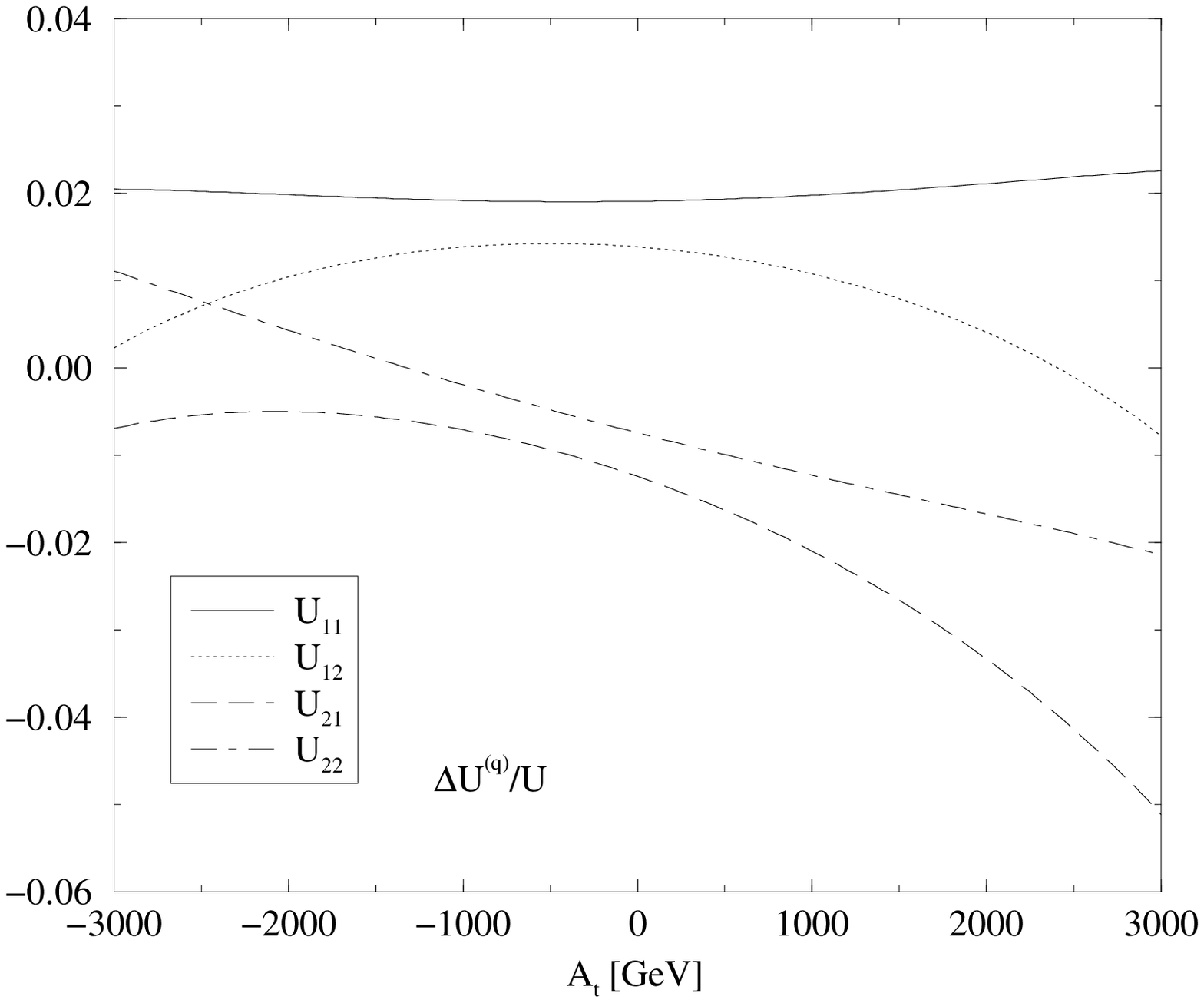}} &
\resizebox{7cm}{!}{\includegraphics{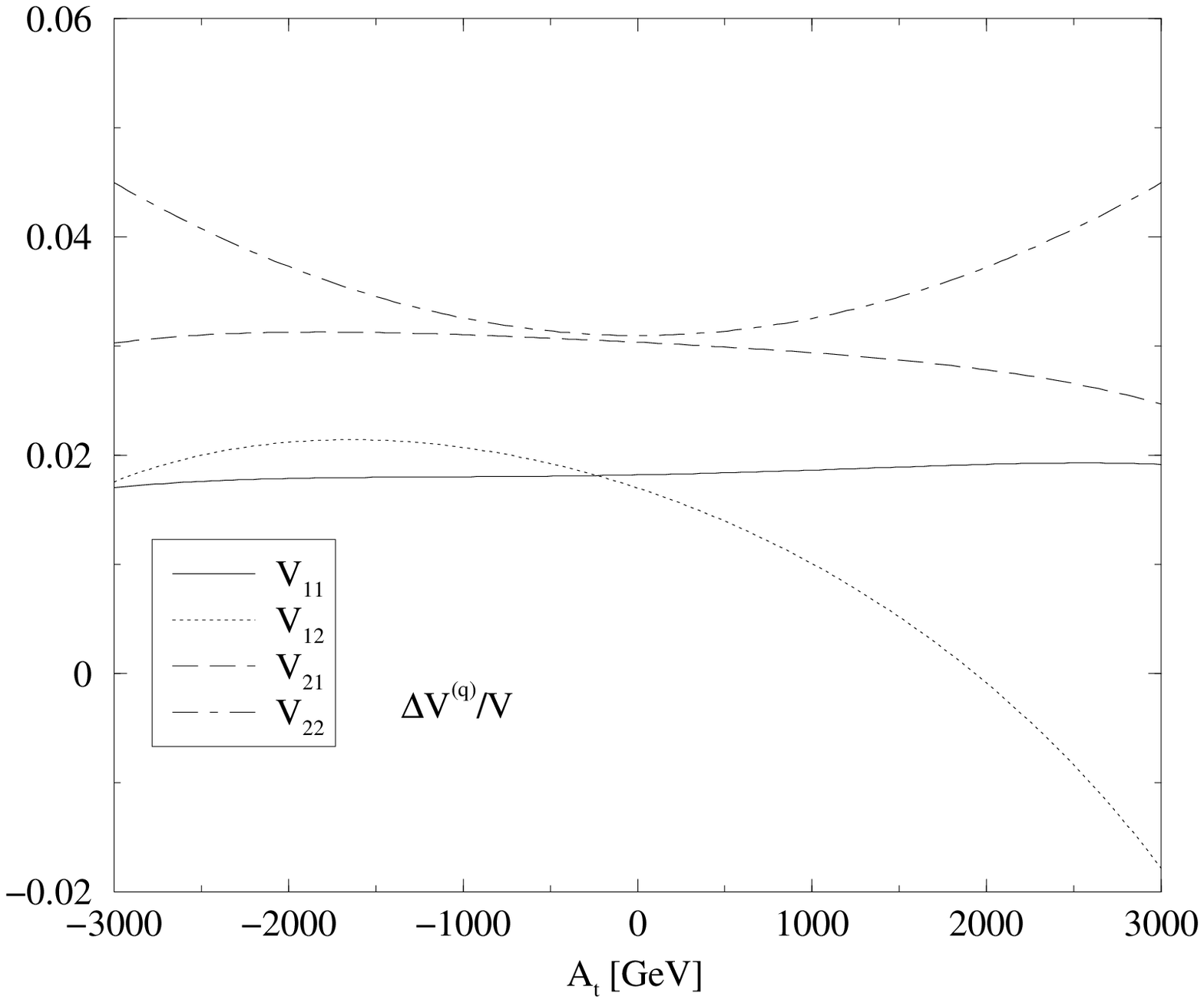}} \\
(a) & (b)
\end{tabular}
\caption{Variation of the chargino effective coupling
  matrices~(\ref{eq:effectivegen}) as a function of the top-squark
  soft-SUSY-breaking trilinear coupling $A_t$ for SPS 1a, but with a
  squark mass scale of 1\TeV. \label{fig:UVAt}}
\end{figure}

We are also interested in the variation with the soft-SUSY-breaking squark
trilinear coupling $A_q$. For the first and second generation squarks
the variation is negligible. The corrections show some variation with
$A_b$, but it is well below the 1\% level. In Fig.~\ref{fig:UVAt} we
show the variation of the chargino effective couplings with $A_t$. In
this figure we have chosen a squark mass scale of 1\TeV. Since $A_t$
enters the computation of the physical top-squark masses, choosing 
a \textit{light} squark mass scale ($\sim 500\GeV$) would produce light
physical top-squark masses ($\sim 100\GeV$) for certain values of
$A_t$. In that case one would find large variations in the
corrections which are due to the presence of light top-squark particles,
and not to the trilinear coupling \textit{per se}. Furthermore, these
light top-squark particles could be produced at the LC, and their
properties precisely measured. In this figure we see large variations of
the corrections (up to 4\%), mainly in the \textit{higgsino}
components of the charginos ($U_{i2}$, $V_{i2}$). Therefore these
corrections are mainly relevant for the couplings of third generation
sfermions ($\stau,\sbottom,\stopp$). Again, a precise knowledge of $A_t$
is not necessary to provide a prediction with sufficient precision, but
a rough knowledge of the scale and sign is needed.

\section{Conclusions}

We have computed the complete one-loop EW corrections to the
sfermion partial decay widths into charginos and neutralinos. We have
combined these corrections with the QCD ones, and provided a combined
prediction for these observables. The corrections from the EW sector can
be of the same order as that of the QCD sector, therefore both kinds of
effects must be taken into account on the same footing.

In these corrections non-decoupling effects appear. These effects are due to
two kinds of splittings among the particle masses: a splitting between a
particle and its SUSY partner (given by the soft-SUSY-breaking masses);
and a splitting among the SUSY particles themselves. In this situation
the radiative corrections grow with the logarithm of the largest SUSY
particle of the model. In this scenario some of the particles
(presumably strongly interacting particles) are heavy, and can only be produced
at the LHC, whereas another set of particles (selectrons, lightest
charginos/neutralinos) can be studied at the LC, and their properties
measured with a precision better than 1\%. 

In order to provide a prediction at the same level of accuracy, one
needs a knowledge of the squark masses (and $A_t$) obtained from the LHC
measurements, but a high precision measurement of the squark parameters
is not necessary. 

The effects of squarks can be taken into account by the use of effective
coupling matrices in the chargino/neutralino sector. These effects can
be extracted from LC data, by finding the finite difference between the
mixing matrices obtained from the chargino/neutralino masses, and the
mixing matrices obtained from the couplings analysis.

Of course, to reach the high level of accuracy needed at the LC the
complete one-loop corrections to the observables under study is needed,
but the effective coupling matrices form a necessary and universal
subset of these corrections.

\vspace{0.2cm}

\noindent{\textbf{Acknowledgments:}}
This collaboration is part of the network ``Physics at Colliders'' of the
European Union under contract HPRN-CT-2000-00149.
The work of J.S.
has been supported in part by MECYT and FEDER under project FPA2001-3598.

\providecommand{\href}[2]{#2}

\end{document}